\documentclass[12pt]{article}

\usepackage{epsfig}
\usepackage[figuresright]{rotating}
\begin{document}
\newcommand{\vet}        {\overrightarrow}
\newcommand{\dafne}      {DA$\Phi$NE }
\newcommand{\repsilon}   {\Re(\epsilon'/\epsilon)}
\newcommand{\vud}        {V_{ud}}
\newcommand{\vus}        {V_{us}}
\newcommand{\vub}        {V_{ub}}
\newcommand{\kpidue}     {K_{\pi2}}
\newcommand{\kmudue}     {K_{\mu2}}
\newcommand{\kmutre}     {K_{\mu3}}
\newcommand{\ketre}      {K_{e3}}
\newcommand{\pimudue}    {\pi_{\mu2}}
\newcommand{\kmu}        {K_{\mu}}
\newcommand{\kpi}        {K_{\pi\pi^0}}

\newcommand{\kls}        {K_S^0K_L^0}
\newcommand{\kpm}        {K^+K^-}

\newcommand{\phikls}     {\phi \rightarrow K_S^0K_L^0}
\newcommand{\phikpm}     {\phi \rightarrow K^+K^-}
\newcommand{\phippp}     {\phi \rightarrow \pi^+ \pi^- \pi^0}
\newcommand{\eeteeg}     {e^+e^-\rightarrow e^+e^-\gamma}

\newcommand{\kmunu}      {K \rightarrow \mu \nu}
\newcommand{\kmunug}     {K \rightarrow \mu \nu (\gamma)}
\newcommand{\kpipi}      {K \rightarrow \pi \pi^0}
\newcommand{\pimunu}     {\pi \rightarrow \mu \nu}
\newcommand{\pigg}      {\pi^0 \rightarrow \gamma \gamma}
\newcommand{\pimunug}    {\pi \rightarrow \mu \nu (\gamma)}

\newcommand{\kpmunu}     {K^{+} \rightarrow \mu^{+}\nu}
\newcommand{\kpmunug}    {K^{+} \rightarrow \mu^{+}\nu\gamma}
\newcommand{\kpmunugall} {K^{+} \rightarrow \mu^{+}\nu(\gamma)}
\newcommand{\kppipi}     {K^{+} \rightarrow \pi^{+}\pi^0}
\newcommand{\kppipigall} {K^{+} \rightarrow \pi^{+}\pi^0(\gamma)}
\newcommand{\kpxpi}      {K^{+} \rightarrow X^{+}\pi^0}
\newcommand{\kppimu}     {K^{+} \rightarrow \pi^0\mu^{+}\nu}
\newcommand{\kppie}      {K^{+} \rightarrow \pi^0 e^{+}\nu}
\newcommand{\kppil}      {K^{+} \rightarrow \pi^0 l^{+}\nu}
\newcommand{\kptaup}     {K^{+} \rightarrow \pi^{+}\pi^0 \pi^0}
\newcommand{\kpppp}      {K^{+} \rightarrow \pi^{+}\pi^{+}\pi^{-}}

\newcommand{\kmmunu}     {K^{-} \rightarrow \mu^{-}\overline{\nu}}
\newcommand{\kmmunug}    {K^{-} \rightarrow \mu^{-}\overline{\nu}\gamma}
\newcommand{\kmmunugall} {K^{-} \rightarrow \mu^{-}\overline{\nu}(\gamma)}
\newcommand{\kmpipigall} {K^{-} \rightarrow \pi^{-}\pi^0(\gamma)}
\newcommand{\kpmmunu}    {K^{\pm} \rightarrow \mu^{\pm}\nu}
\newcommand{\kpmmunug}   {K^{\pm} \rightarrow \mu^{\pm}\nu\gamma}
\newcommand{\kpmmunugall}{K^{\pm} \rightarrow \mu^{\pm}\nu(\gamma)}
\newcommand{\kpmpipi}    {K^{\pm} \rightarrow \pi^{\pm}\pi^0}
\newcommand{\kpmpimu}    {K^{\pm} \rightarrow \pi^0\mu^{\pm}\nu}
\newcommand{\kpmpie}     {K^{\pm} \rightarrow \pi^0 e^{\pm}\nu}
\newcommand{\kpmtaup}    {K^{\pm} \rightarrow \pi^{\pm}\pi^0 \pi^0}
\newcommand{\kpmppp}     {K^{\pm} \rightarrow \pi^{\pm}\pi^{+}\pi^{-}}

\newcommand{\klppp}      {K_L \rightarrow \pi^+ \pi^- \pi^0}

\newcommand{\tz}         {$T0$}

\title{The measurement of the absolute branching ratio of the $\kppipigall$ decay at KLOE.}
\date{KLOE Collaboration:}
\maketitle
\author{
F.~Ambrosino,
A.~Antonelli, 
M.~Antonelli, 
F.~Archilli,
C.~Bacci,
P.~Beltrame,
G.~Bencivenni, 
S.~Bertolucci, 
C.~Bini, 
C.~Bloise, 
S.~Bocchetta, 
V.~Bocci,
F.~Bossi,
P.~Branchini,
R.~Caloi,
P.~Campana, 
G.~Capon, 
T.~Capussela,
F.~Ceradini,
S.~Chi,
G.~Chiefari, 
P.~Ciambrone,
E.~De~Lucia,
A.~De~Santis, 
P.~De~Simone, 
G.~De~Zorzi,
A.~Denig,
A.~Di~Domenico,
C.~Di~Donato,
S.~Di~Falco,
B.~Di~Micco,
A.~Doria,
M.~Dreucci,
G.~Felici, 
A.~Ferrari,
M.~L.~Ferrer, 
G.~Finocchiaro,
S.~Fiore,
C.~Forti,       
P.~Franzini,
C.~Gatti,      
P.~Gauzzi,
S.~Giovannella,
E.~Gorini, 
E.~Graziani,
M.~Incagli,
W.~Kluge,
V.~Kulikov,
F.~Lacava, 
G.~Lanfranchi, 
J.~Lee-Franzini,
D.~Leone,
M.~Martini,
P.~Massarotti,
W.~Mei,
S.~Meola,
S.~Miscetti, 
M.~Moulson,
S.~M\"uller,
F.~Murtas, 
M.~Napolitano,
F.~Nguyen,
M.~Palutan,          
E.~Pasqualucci,
A.~Passeri,  
V.~Patera,
F.~Perfetto,
M.~Primavera,
P.~Santangelo,
G.~Saracino,
B.~Sciascia,
A.~Sciubba,
F.~Scuri, 
I.~Sfiligoi,     
T.~Spadaro,
M.~Testa,
L.~Tortora, 
P.~Valente,
B.~Valeriani,
G.~Venanzoni,
R.~Versaci,
G.~Xu.
}
\vskip 1cm
\begin{center}
  \begin{abstract}
    \noindent
    The preliminary result on the absolute branching ratio of the decay $\kppipigall$, obtained by the 
    KLOE experiment operating at the DA$\Phi$NE Frascati $\phi$-factory, is 
    presented.
  \end{abstract}
\end{center}

\section{DA$\Phi$NE and KLOE}
\noindent 
The DA$\Phi$NE e$^+$e$^-$ collider operates at a total energy 
$W=1020$ MeV, the mass of the $\phi$(1020)-meson.
Since 2001, the KLOE experiment has collected an integrated luminosity of 
about 2.5 fb$^{-1}$.
Results presented below are based on an integrated luminosity of about 250~pb$^{-1}$.
The KLOE detector consists of a large cylindrical drift chamber, DC, surrounded
by a lead/scintillating-fiber electromagnetic calorimeter, EMC. 
The drift chamber \cite{bib:dc} is 4~m in diameter and 3.3~m long,
has full stereo geometry and operates with a 90$\%$ helium - 10$\%$ isobutane gas mixture.
The momentum resolution is $\sigma(p_{T})/p_{T} \sim 0.4\%$. 
Two track vertices are reconstructed with $\sim$ 3 mm resolution. 
The calorimeter \cite{bib:emc}, composed of a barrel and two endcaps,
covers 98\% of the solid angle. 
Energy and time resolution are $\sigma(E)/E = 5.7\%/\sqrt{E({\mbox GeV})}$ and
$\sigma(t) = 57 {\mbox ps}/ \sqrt{E({\mbox GeV})} \oplus 100 {\mbox ps}$.
A superconducting coil surrounds the detector and provides a solenoidal field
of 0.52~T.
The KLOE trigger \cite{bib:trg}, uses calorimeter and drift chamber
information. 
For the present analysis, only events triggered by the calorimeter have been used. 
\section{The tag mechanism}
The $\phi$-meson decays most of the times into $K\bar{K}$ pairs;
these are quasi anti-collinear in the laboratory due to the small crossing angle of the e$^+$e$^-$ beams.
Therefore the detection of a $K(\bar K)$ guaranties the presence of a  
$\bar K (K)$ of given momentum and direction. 
%
%
Therefore identified $K^{\mp}$ decays tag a $K^{\pm}$ beam and provide 
the normalization sample for signal count.
This procedure is a unique feature of a $\phi$-factory and gives the possibility
of measuring absolute branching ratios. 
Charged kaons are tagged using their two-body decays, 
$K^{\pm}\rightarrow \mu^{\pm} \nu_{\mu}$ and
$K^{\pm}\rightarrow \pi^{\pm} \pi^0$, accounting for 
$\sim$85\% of the total decay channels.
We have about  $1.5 \times 10^6 K^+K^-$ events/pb$^{-1}$. 
The two body decays are identified as peaks in the momentum spectrum of the
charged decay particle evaluated in the kaon rest frame and assuming the pion mass.
\section{Measurement of the absolute branching ratio of the $\kppipigall$ decay.}
\noindent
The measurement of the branching ratio (BR) is performed using 250 pb$^{-1}$ collected at $\phi$ peak.
The normalization sample is given by $\kmmunugall$ tagged events, providing a pure $K^+$ beam
for signal search. 
In order to minimize the impact of the trigger efficiency on the reconstruction
of the signal decay channel, we require the tagging kaon alone to satisfy the 
EMC trigger request, hereafter {\it self-triggering tag}.
Nevertheless a residual dependency of the tagging criteria on the decay mode of the signal kaon
is still present and it is accounted for in the final branching ratio measurement.The decision of 
using $K^-$ to tag and $K^+$ for signal search has been taken to neglect corrections to the BR from 
nuclear interactions (NI) of the kaon ($\sigma_{NI}(K^+) \sim \sigma_{NI}(K^-) / 10^{2}$). The choice
of $K^-_{\mu 2}$ decays for tagging purpouses allows us to separate as much as possible the tag hemisphere 
from the signal hemisphere, minimizing possible interference in track reconstruction
and cluster association. \\

\noindent
The signal selection of $\kppipigall$ decays uses DC information only. The $K^+$ is 
identified by a positive track  moving outwards in the DC with momentum
$70 < p_K < 130$ MeV/c and having point of closest approach (PCA) to the interaction point
 with $\sqrt{x^2_{PCA} +y^2_{PCA}} < 10$ cm and $|z_{PCA}| < 20$ cm.
Decay vertices (V) in the drift chamber fiducial volume 
are selected, 40 $< \sqrt{x^2_{V} + y^2_{V}}<$ 150 cm, 
with the momentum difference between the kaon and the secondary 
 $-320 < \Delta p = | \vet p_K | - | \vet p_{sec} | < -50$ MeV/c and 
the charged decay particle momentum in the kaon rest frame in pion mass hypothesis
$50 < p^{\ast} < 370$ MeV/c. \\
\begin{figure}[!bhtp]
   \begin{center}
   \epsfig{file=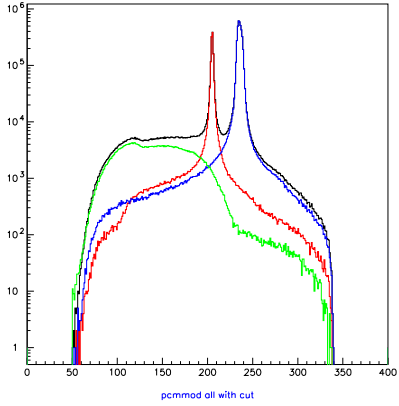,width=6.cm}
   \end{center}
   \caption{Spectra of the charged secondary momentum in the kaon mass
            rest frame assuming the pion mass obtained from MonteCarlo simulation. 
            Two peaks are visible, the {\it $K_{\pi 2}$ peak} at $205$
            MeV and the {\it $K_{\mu 2}$ peak} at $236$ MeV.
            Blue corresponds to $\kpmunu$, red to
            $\kppipi$ and green to $three-body$ decays.}
   \label{fig:eventsel}
\end{figure}

\noindent
After this selection, the $\kppipigall$ signal count is extracted from the fit
of the $p^{\ast}$ distribution in the window starting from $p^{\ast}_{cut}$=180 MeV/c (see figure \ref{fig:eventsel}). 
This spectrum exhibits two peaks, the first at about $236$ MeV/c from $\kpmunu$ decays, {\it $K_{\mu 2}$ peak},
and the second at about $205$ MeV/c from $\kppipi$ decays, {\it $K_{\pi 2}$ peak};
lower $p^{\ast}$ values are due to three body decays. 
The momenta of the charged secondaries produced in the kaon decay 
have been evaluated in the kaon rest frame using the pion mass hypothesis.
Therefore the {\it $K_{\pi 2}$ peak}, evaluated using the correct mass hypothesis, 
appears to be symmetric while the {\it $K_{\mu 2}$ peak} is asymmetric do to the incorrect mass 
hypothesis used (pion instead of muon). \\

\begin{figure}[hptb]
   \begin{center}
   \epsfig{file=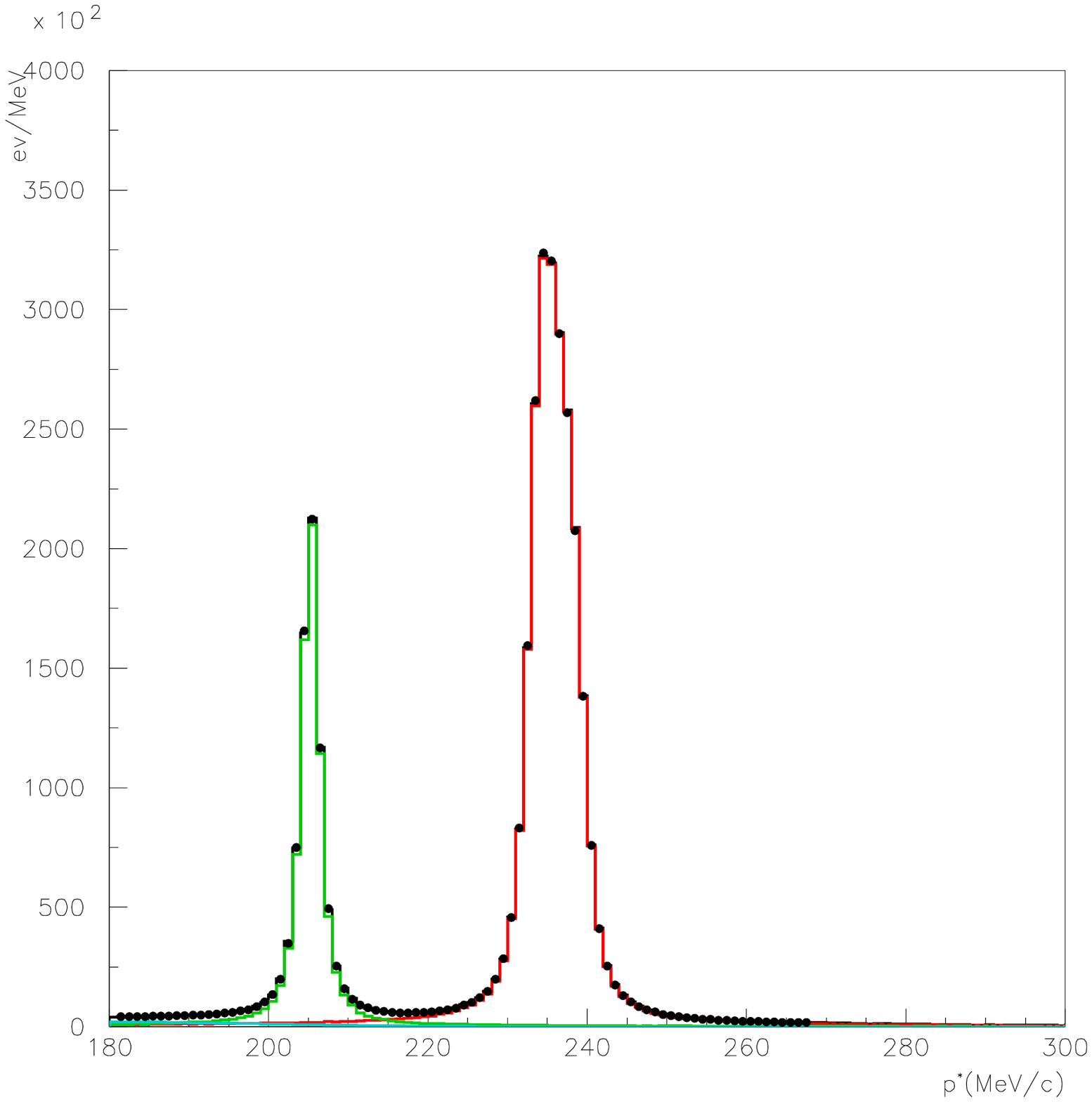,width=6cm}
   \epsfig{file=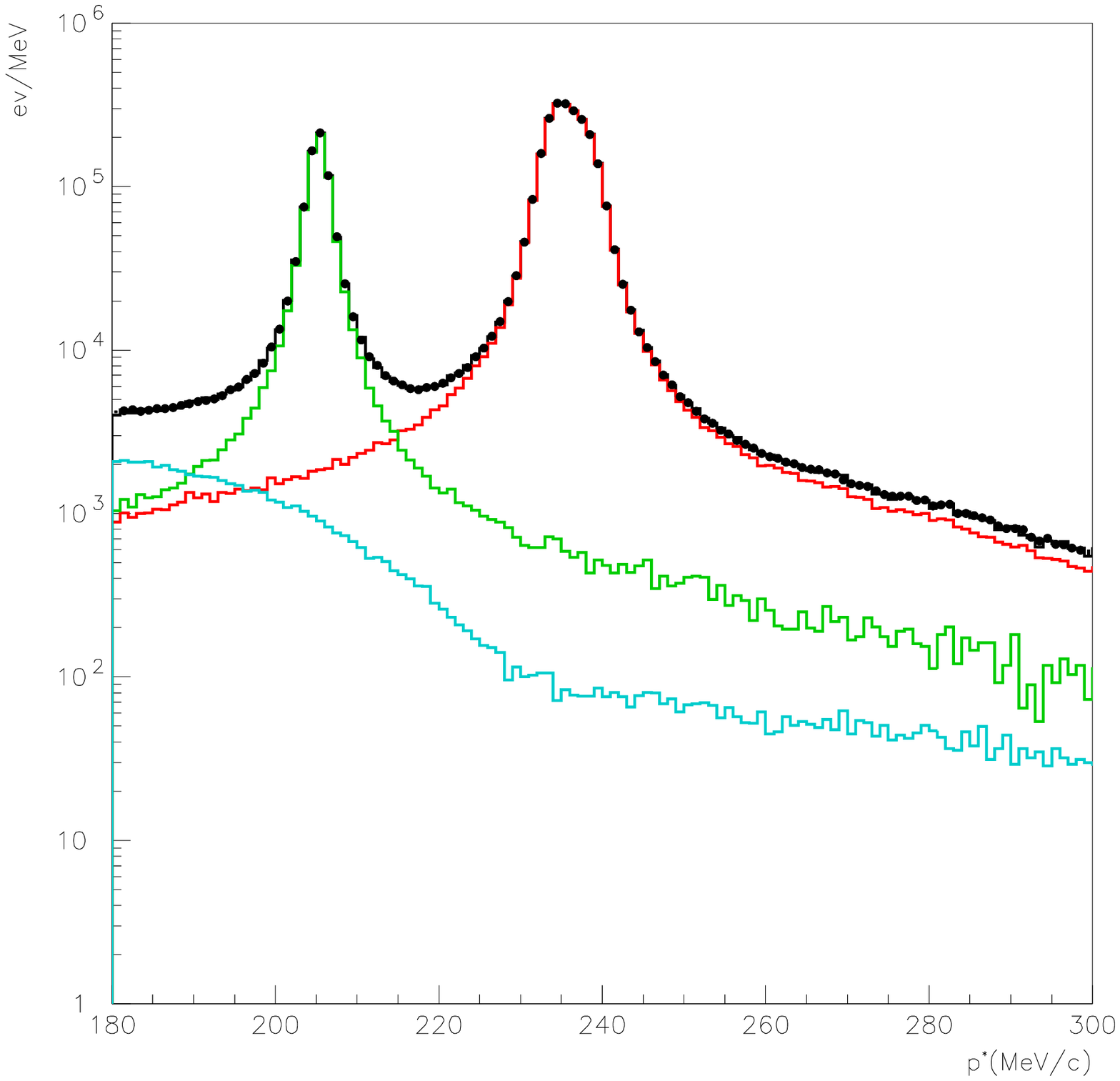,width=6cm}
   \end{center}
   \caption{Fit of the $p^\ast$ distribution: black dots are DATA to be fit and 
            solid black line is the fit output. The three contributions used to fit the DATA
            are: $K_{\mu 2}$ peak (red), $K_{\pi 2}$ peak (green) and $three-body$ decays (light blue).}
   \label{fig:fit_res}
\end{figure}

\noindent
The fit to the $p^{\ast}$ distribution is done using the following three contributions:
\begin{enumerate}
\item {\it $K_{\mu 2}$ peak}: this contribution accounts for $\kpmunugall$ decays and 
                      it is taken directly from a DATA control sample 
                      selected using calorimetric information only;
\item {\it $K_{\pi 2}$ peak}: this contribution accounts for $\kppipigall$ decays and 
                      it is taken directly from a DATA control sample 
                      selected using calorimetric information only;
\item $three-body$ decays: this contribution accounts for $three-body$ decays and it is taken from MC simulation.
\end{enumerate}
Figure \ref{fig:fit_res} shows the result of the fit of the
$p^\ast$ distribution performed on the selected DATA sample.
Using a total number of
tagging events $N_{tag}=12,113,686$ and $p^{\ast}_{cut}=180$ MeV/c, 
we obtain $N_{\kppipigall}|_{FIT}=818,347 \pm 1,912$.
Different colours indicate the different contributions: green 
for $\kppipigall$ decays, red for  $\kpmunugall$
decays and light blue for $three-body$ decays. \\

\noindent
The efficiency has been evaluated directly on DATA from
a control sample selected using calorimetric information only,
to avoid correlation with the DC driven sample selection.
Given the tag by $\kmmunugall$ decays, 
the control sample selection for the evaluation of this efficiency
is given by $\kpxpi$ decays identified via the reconstruction of  $\pigg$ decay vertices.
Corrections to the efficiency accounting for possible distortions induced by the
control sample selection have been evaluated using MC simulation. \\

\noindent
A preliminary evaluation of systematic uncertainties has been done and
the preliminary measurement of the absolute branching ratio of the $\kppipigall$ decay, at the few permil
level of precision, will be presented at the conference.

\end{document}